\documentclass[12pt]{article}
\usepackage{graphicx}
\usepackage{epsfig}
\usepackage{graphicx}
\usepackage{amsmath}
\usepackage{hhline}
\usepackage{amssymb}
\usepackage{times}
\usepackage{cite}

\newlength{\dinwidth}
\newlength{\dinmargin}
\title{The contribution of cosmic rays to global warming. }
\author{T. Sloan$^{1}$ and A. W. Wolfendale $^{2}$\\
$(1)$ Department of Physics, Lancaster University, UK\\
$(2)$ Department of Physics, Durham University,\\ Durham, UK}

\begin{document}
\maketitle

\begin{abstract}
A search has been made for a contribution of the  
changing cosmic ray intensity to the global warming  
observed in the last century. The cosmic ray intensity shows 
a strong 11 year cycle due to solar modulation and the overall rate 
has decreased since 1900.  These changes in cosmic ray intensity are 
compared to those of the mean global surface temperature to attempt 
to quantify any link between the two. 
It is shown that, if such a link exists, the changing cosmic ray intensity 
contributes less than 8\% to the increase in the mean global 
surface temperature observed since 1900. 

\end{abstract}

\section{Introduction}

There are many claims that cosmic rays (CR) 
influence the climate significantly, so that some of the Global Warming (GW) 
seen since industrialisation could be due to the effects of them  
(e.g.see Dorman 2009, Rao 2011, Veizer 2005, Lee 2003, Weber 2010). 
The essence of these works is that because the CR  
intensity has been falling over the last century and that CR could 
influence cloud formation  
(Svensmark and Friis-Christensen 1997, Marsh and Svensmark 2000, 
Svensmark 2007, Svensmark and Calder 2007), 
a significant contribution to GW could result. 

The formation of condensation nuclei of nanometer sizes, on which clouds 
seed, is not well understood. The standard picture is that such nuclei 
form on atmospheric impurities such as sulphates, microscopic dust or 
salt particles (Taylor 2005).  A further contribution is possible, namely, 
through atmospheric ionization as has been shown theoretically by  Yu and Yu 
and Turco (Yu 2002) and experimentally by Enghoff et al 
(Enghoff et al. 2011). The crucial question 
is the extent to which ionization from cosmic rays affects the rate of 
formation of condensation nuclei compared to that provided by the 
standard known processes.  If the effect of CR is significant it will  
influence GW.  

In order to give a quantitative answer to the question of the contribution 
of the changing CR rate to GW we compare the 
variation of the galactic CR intensity over the last century to the 
observed mean global surface temperature. 
Previous studies have been made of correlations between various 
climatological processes and the mean global surface temperature 
(e.g. Lean and Rind 2008). However, CR have not been explicitly considered 
and this question is addressed here.  Firstly, the long term trend 
of the CR rate and the global temperature 
are compared. Secondly, a search is made for an 11 year cycle in the 
temperature data of a comparable shape to that caused by the solar 
modulation of the CR intensity.

\section{The long term trends of the cosmic ray intensity and temperature}

Figure \ref{fig1} (upper plot) shows the variation of the galactic 
CR rate in terms of the equivalent Climax neutron monitor rate. The Climax 
neutron monitor rate is here used as a proxy for the rate of 
production of atmospheric ionization by galactic CR.  The neutron monitor 
rates have been derived from 1428 to 2005 (McCracken and Beer 2007) 
directly from the Climax 
neutron monitor record (1953-2006), from ion chamber data (1933-65) 
and from the observations of $^{10}$Be in ice cores.
The lower plot in figure \ref{fig1} shows the mean global surface 
temperature anomaly 
as a function of time (NASA 2011). The CR rate has fallen with time 
since 1880 as the global temperature has risen. This has been used 
to postulate that cosmic rays play a significant role in global 
warming (Rao 2011, Svensmark 2007). A clear 11 year modulation 
due to solar activity 
can also be seen in the variation of the CR rate in the upper plot of figure 
\ref{fig1}.  

Figure \ref{fig2} shows the data from figure \ref{fig1} with 11 year smoothing 
applied to illustrate the long term trends (removing the effects of the solar 
cycle). It can be seen that the CR rate decreased rapidly from 1880 
to 1950 when the temperature rise was relatively small. On the other 
hand the CR rate changed little from 1950-2000 when the temperature 
rise was bigger (Lockwood and Fr\"ohlich 2007). Hence the variation 
of the temperature 
with time matches poorly the variation of the CR rate with time. 
Such a match would be 
expected if CR played a significant role in global warming.    
It can be seen that the CR rate decreased in intensity by 
$\sim$16\% from 1880 to 1952 during which time the mean global 
temperature rose 
by $\sim$0.3$^\circ$ while from 1952 to 2000 the CR rate decreased  
by $\sim$3\% and the temperature rose by $\sim$0.6$^\circ$. 


Figure 3 shows the values of the temperature anomalies $\Delta T$, 
plotted 
as a function of CR rate, each taken from figure 2. It can be seen that in 
the early part of the 20th century there was a roughly linear relationship 
between the two. Linear fits were made to the data in various time ranges 
starting from 1886. The slopes, d$\Delta$T/dCR, varied in values from 
-0.008 to -0.022 $^\circ$C per \% change in the Climax NM rate as 
the end year of the fit varies from 
the year 1930 to 2000. Cycling through each end year in this time range, 
the mean value of d$\Delta$T/dCR was 
-0.016$^\circ$ C per \% change in the Climax NM rate with root mean square 
deviation of the readings of 0.003.  If there is a causal link between 
CR and the global temperature, this value of d$\Delta$T/dCR implies  
that the 3\% change in CR rate observed since 1952 would give a temperature 
change of $\sim$0.05$^\circ$C. The observed 
change in temperature since 1952 is $\sim$0.6$^\circ$C. Hence less than 
$\sim$8\% of the observed rise in temperature since 1952 can be ascribed 
to CR. We take this as an upper limit on the contribution of CR to GW. 


\section{The 11 year cycle}
\label{sec3}

\begin{figure}[htb]
\includegraphics[width=40pc]{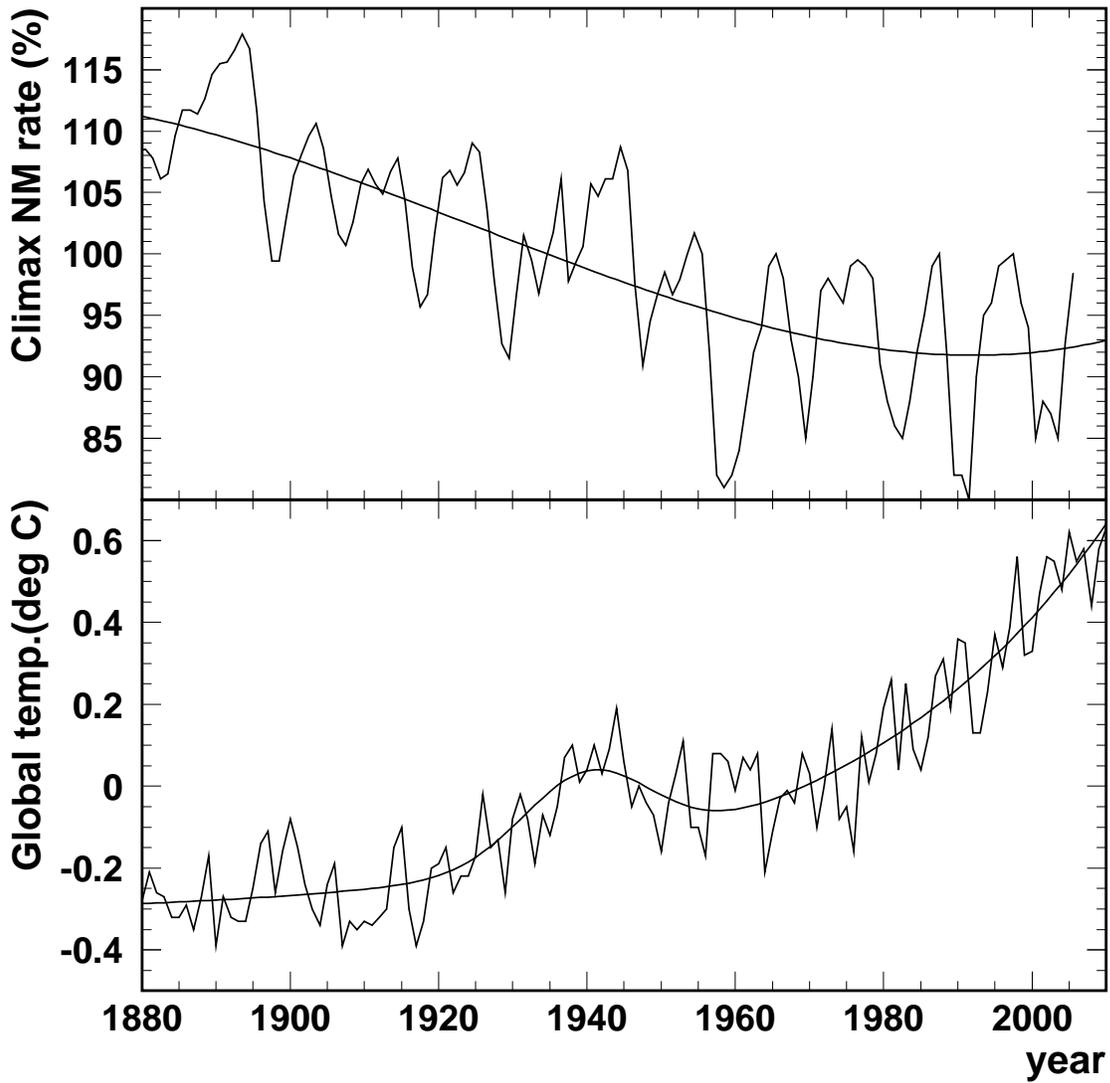}
\caption{\label{fig1} {\bf Upper panel} shows the equivalent Climax 
neutron monitor rate (McCracken and Beer 2007) against time. 
{\bf Lower panel} shows the mean global surface temperature  
from the GISS data against time. Both data sets are yearly averages. 
The smooth curves show the fitted functions used to detrend the data 
(see section \ref{sec3}).  } 
\end{figure}

\begin{figure}[htb]
\includegraphics[width=40pc]{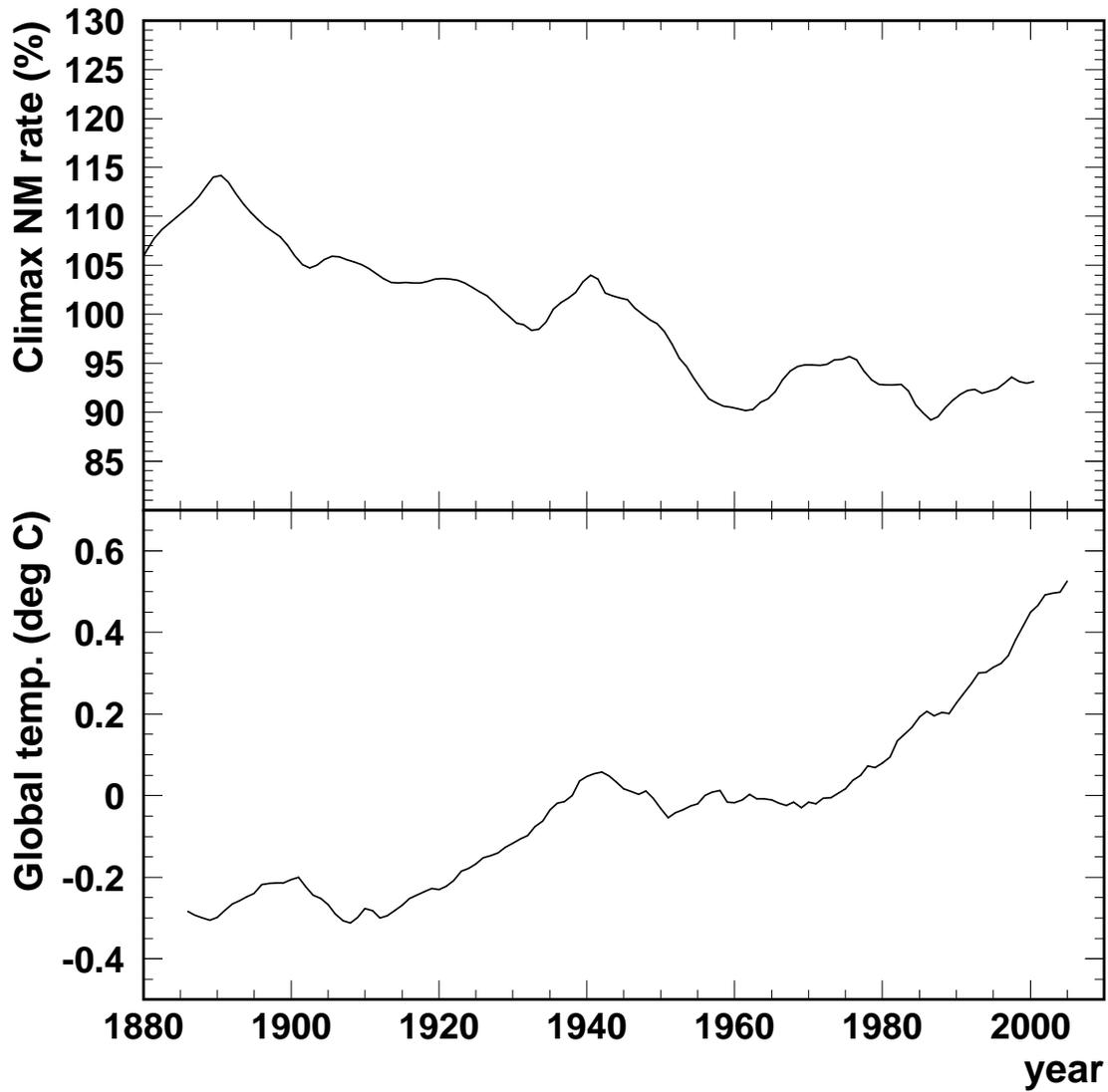}
\caption{\label{fig2} {\bf Upper panel} shows the equivalent Climax 
neutron monitor rate (McCracken and Beer 2007) against time with 
11 year smoothing 
to illustrate the trend and smooth out the effects of the solar cycle. 
{\bf Lower panel} shows the mean global surface temperature  
from the GISS data against time with the same 11 year smoothing applied. } 
\end{figure}

The CR rates in figure \ref{fig1} show a clear 11 year cyclic wave 
due to solar activity whilst such a wave is difficult to discern  
in the temperature variation. A Fourier analysis was made of each 
of the variations in figure 
\ref{fig1} in order to measure wave amplitudes. The Fourier 
amplitudes, $a$, at angular frequency $\omega$ were derived from the integral 
 \begin{equation}
a=\frac{2}{NT} \int_0^{NT} exp(i\omega t) f(t) dt= \frac{2}{NT} \sum_{bins}exp(i\omega t) f(t) \delta t
\label{fourier}
\end{equation}  
where $T=2 \pi/\omega$ is the period at this frequency, $N$ is the 
number of periods of data analysed, $f(t)$ is the time distribution in bins 
of width $\delta t$=1 year. Note that the values of $a$ 
in equation \ref{fourier} are 
absolute amplitudes in the same units as the data. These 
amplitudes will be used throughout 
this work rather than the more conventional power spectra. There were no 
missing years in any of the time series. To avoid generating spurious 
peaks only whole numbers of periods 
were analysed. This was done in two stages:   
first from the start of the series and 
second up to the end of the series, averaging the amplitudes from each  
stage. In order to reduce noise proportional to the 
reciprocal of frequency, the data were detrended by fitting simple 
functions. For the equivalent Climax data a 
cubic polynomial was fitted and subtracted. For the temperature data 
an exponential function of time was employed together with a Gaussian shape 
to account for the anomalous temperature peak at 1942. 
Optimum fits to the data were obtained using an exponential with time 
constant 36.5 years and a Gaussian shape centred at 1942 with amplitude 
0.21$^\circ$C and standard deviation 8.67 years. The trend curves are 
compared with the data in figure \ref{fig1}.  Fourier analysis 
was then carried out on the data with these trend curves subtracted. 

Figure \ref{fig4} shows the results of the Fourier analysis. 
The solid curve shows the components of the Fourier amplitudes 
of the temperature data in phase (positive amplitude) or 
antiphase (negative amplitude) to those  
measured from an analysis of the equivalent 
Climax rates (shown in figure \ref{fig1}). To test the sensitivity 
to the Climax shape the detrended Climax data from figure \ref{fig1},  
normalised to temperature waves of different amplitudes,  
were added to the detrended GISS temperature series (NASA 2011) and 
the analysis repeated. 
The dotted and dashed curves in figure \ref{fig4} show the effects 
of adding waves normalised to amplitudes of either plus (upper panel)  
or minus (lower panel) 0.035$^\circ$ and 0.07$^\circ$, respectively. 
It can be seen from the solid curves in figure \ref{fig4} that we do 
not see a significant 
shape similar to the equivalent Climax data in the measured amplitudes. 
This indicates that the amplitude of the 11 year cycle on the global 
temperatures is small.    

To assess the significance of the structures seen in the solid curve 
in figure 
\ref{fig4} randomly generated spectra were passed through 
the Fourier analysis programs. 
These were generated with the same root mean square (RMS) 
noise  about the trend curve as measured from the data. The result of 
analysing 1000 random spectra showed that for periods 
between 5 and 15 years the RMS deviation of the Fourier amplitudes about 
zero was 0.0124$\pm$0.0001 $^\circ$C. This is to be compared 
to the value of 0.0115$\pm$ 0.0023 $^\circ$C 
observed from the temperature data. These two numbers are equal within 
the uncertainties showing that the fluctuations in the data 
are mainly due to noise. 

The largest peak observed in the temperature data shown by the solid 
curves in figure \ref{fig4} has   
amplitude -0.04$^\circ$C and occurs at period 
7.5 years. Peaks of the size of this one or greater occurred in 
14\% of the random spectra, so this peak is most probably 
due to noise. Hence  
no significant structure could be found in the temperature data.      
In particular, there was no significant structure 
near to that expected from the CR signal at periods 
between 10-11 years.   
 
We proceed to assess the minimum detectable Climax-like signal. 
If a Climax-like signal exists in the temperature data in figure 
\ref{fig1}, adding in antiphase a normalised measured Climax 
spectrum  will reduce the RMS value of the Fourier amplitudes. 
The Climax signal necessary to minimise the RMS value of the Fourier 
amplitudes between periods from 3-12 years in the solid curve of 
figure \ref{fig4} was found to be equivalent to a temperature 
amplitude of -0.023$^\circ$C. This is taken as the measured temperature 
wave amplitude although such a small signal cannot be assumed to be 
real since such values are frequently generated by the random spectra.  
Climax waves normalised to temperatures of different amplitudes
were then added to the random spectra. It was found that, for 
the addition of a Climax wave normalised to a temperature amplitude 
of -0.047$^\circ$C, 10\% of the random spectra would have been given 
a temperature reading closer to zero than the measured value 
of -0.023$^\circ$C i.e. a temperature wave 
of amplitude -0.047$^\circ$C has a 10\% probability to be masked by noise.  
Hence the minimum detectable temperature amplitude is -0.047$^\circ$C
at 90\% confidence level.  
(or 0.094$^\circ$C peak to peak). An analysis using the absolute 
amplitudes without reference to the Climax phase gave a 
somewhat larger 90\% confidence 
limit on the amplitude of the temperature wave of -0.06$^\circ$C 
(or 0.12$^\circ$C peak to peak).  
A similar analysis of the positive Climax signals using the 
phase information shows that 
waves of amplitude 0.028$^\circ$C would have been missed in 10\% 
of the cases. Hence the upper limit on the positive temperature 
amplitude is this value at 90\% confidence level.

The average peak 
to peak amplitude of the 11 year wave on the Climax rate due to solar 
modulation is 14.7\% during 1900-2005. Any link between CR and clouds 
will cause an antiphase 11 year wave on the temperature. This has been 
shown above to have amplitude of less than 0.094$^\circ$C peak to peak. 
Hence  d$\Delta$T/dCR is less than 0.0064$^\circ$C per \% change in 
the Climax rate at 90\% confidence level.  
The observed temperature rise from 1900-2005 is 0.8$^\circ$ and 
the total change in the Climax NM rate was 11\% (figures 
\ref{fig1} and \ref{fig2}). Hence the total change in global 
temperature due to CR 
must be less than 0.07$^\circ$C from 1900-2005 which is less than 
9\% of the observed GW at 90\% confidence level. This figure would rise to 
11\% of the GW if the value of d$\Delta$T/dCR without the Climax 
phase information is used. 
 

\section{The effect of time constants or delays}

The variation of the CR rate could be made compatible 
with that of the global temperatures by the  
introduction of either an integrating time constant or a delay. 

The effect of a delay is illustrated 
in figure \ref{fig5} which shows the global mean temperature anomaly 
(Brohan et al., 2006) 
\footnote{The CRU data were used here in order to extend the time series} 
as a function of time compared to scaled values of the 
atmospheric concentration of carbon dioxide (the main green house gas) 
and those of the Climax rates. It can be seen from the upper plot in 
figure 5 that if a delay of 35 years is introduced into the CR rates 
the change beginning in 1940 would resemble the observed temperature 
rise in the late 20th century.  A 35 year timescale could probably be 
found somewhere in the ocean system or even geomagnetically through 
mantle convection. However, a memory effect of that duration would 
require the stimulus to be tuned to a very particular regional ocean 
over-turning. This seems implausible.  

The effect of an integrating time constant could also modify 
the CR rate so that it resembles the temperature variation. 
This effect is illustrated 
in the lower plot of figure \ref{fig5} where the CR rates and the 
CO$_2$ concentrations are each 
passed through an integrating time constant of 50 years. Such a 
long time constant is necessary to make the change in CR rate 
resemble the variation of the temperatures. It can be seen from 
figure \ref{fig5} that the CO$_2$ 
concentration gives a much closer representation of the temperature 
variation either with or without a time constant than the CR rates.  

The effect of a 50 year time constant on the global temperature 
would be to filter out the influence  
of the 11 year solar cycle. However, such an effect would also  
filter out other well known short term variations in the 
global temperature such as summer to winter 
changes or day-night variations. Whilst there are effects which 
introduce delayed warming, 
such as the thermal inertia of the oceans, the effects of CR on 
clouds should be on a much shorter 
time scale than the 50 year time constant necessary to make the CR rates
resemble the temperature data. Since this is the main 
mechanism proposed 
for CR to affect the climate, the limits for the contribution of 
CR to GW which we have decribed in 
the previous sections seem safe.  

\section{Discussion of the results}

The contribution of cosmic rays to the increase in the mean surface 
temperature of the Earth during the last century has been 
assessed by comparing the shapes of the variation in the cosmic 
ray rate to those of the observed mean global surface temperatures. 
The long term variation gives an upper limit 
on the contribution of cosmic rays to GW of less than 8\% of the 
observed increase in temperature. It is difficult to gauge the 
statistical confidence level of this value since the change in cosmic 
ray rate is rather uncertain and depends on the precise start date. 
The result is conservative since if we had chosen the start date to 
be 1955 rather than 1952 the change in cosmic ray rate would have 
been zero giving a limit of 0\%. 

The contribution from the shorter term 11 year variation is shown to 
be less than 9\% of the GW since 1900 at the 90\% confidence 
level. This limit would increase, due to dilution of the 
11 year cycle, if the dwell time of 
condensation nuclei is a few years as required 
by the analysis of (Weber 2010). 
Our limit is not inconsistent with the value from (Lean and 
Rind 2008) who deduced from a linear regression analysis 
that there was an 11 year wave on the temperature data due to 
solar insolation with a peak to peak amplitude slightly smaller 
than our limit.    

In conclusion, we deduce that cosmic rays play only a minor part in 
the global warming observed in the last century  
(less than 8\% of the rise in temperature). 
Hence standard processes for cloud seeding must be the 
the dominant mechanisms and ionization seeding of clouds can only 
play a minor part. Finally, using the value of the radiative 
forcing of 1.6 W/m$^2$ 
derived by the Intergovernmental Panel on Climate Change (IPCC 2007)
to produce the observed warming in the last 
century, the radiative 
forcing induced by changing cosmic ray rates must be less than 
0.14 W/m$^2$.  

\section{Acknowledgement}  
We thank K. G. McCracken for providing us with the equivalent Climax 
count rates from the analysis of the direct Climax data, the ion chamber
data and the $^{10}$Be ice core data. We also thank the Dr. 
John C. Taylor Charitable Foundation for financial support.

\section{References}

\begin{enumerate}

\item Brohan P., Kennedy J. J.,  Harris I., Tett S.F.B., 
and Jones P.D., 2006, Uncertainty estimates in regional and global 
observed temperature changes: a new dataset from 1850,  
J. Geophysical Research 111, D12106, doi:10.1029/2005JD006548. \\
(available at http://www.cru.uea.ac.uk/cru/data/temperature/). 

\item Dorman, L., 2009, The Role of Space Weather and Cosmic 
Ray Effects in Climate Change, Climate Change: Observed Impacts on 
Planet Earth, Elsevier.

\item Enghoff M.B., Pedersen J.O.P., Uggerh\o j U.I., Paling S.M. and 
Svensmark H., 2011, Aerosol nucleation induced by a high energy 
particle beam, 
Geophysical Research Letters 38, LO9805, doi;10.1029/2011GL047036.  

\item
Hansen, J., Mki. Sato, R. Ruedy, K. Lo, D.W. Lea, and 
M. Medina-Elizade, 2006: Global temperature change,  
Proc. Natl. Acad. Sci., 103, 14288-14293, \\ doi:10.1073/pnas.0606291103.\\
(available at http://data.giss.nasa.gov/gistemp/graphs/)

\item IPCC 2007, 4th Assessment Report of the IPCC 
(WG1 Figure SPM2). 

\item Lean J.L. and Rind D.H., 2008, How natural and anthropogenic 
influences alter global and regional surface temperatures: 1889 to 2006. 
Geophysical Research Letters {\bf 35} L18701,  
DOI:10.1029/2008GL034864.

\item Lee S.H. et al., 2003, Particle formation by ion nucleation 
in the upper troposphere and lower stratosphere, Science {\bf 298}, 172.

\item Lockwood M. and Fr\"olich C., 2007 Recent oppositely directed 
trends in solar climate forcings and the global mean surface air 
temperature, Proc. Roy. Soc. {\bf A463} no. 2086, 2447-2460 doi: 
10.1098/rspa.2007.1880. 

\item
Marsh N. and Svensmark H., 2000, 
Low cloud properties influenced by cosmic rays, 
 Phys. Rev. Letts. {\bf 85}, 5004. 

\item McCracken, K.G. and Beer, J., 2007, Long-term changes 
in the cosmic ray intensity at Earth, 1428 - 2005, Journ. Geophys. Res. 
112, A 10101.

\item NASA 2011, available at http://data.giss.nasa.gov/gistemp/ 

\item NOAA 2011, Tans P. and Keeling R. (Scripps 
Institution of Oceanography).  The CO2 data are available at 
www.esrl.noaa.gov/gmd/ccgg/trends/  

\item Rao, U.R., 2011, Contributions of changing galactic 
cosmic ray flux to global warming, Current Science, 100, 2, 223.

\item Svensmark H. and Friis-Christensen E., 1997,  
Variation of cosmic ray flux and global cloud coverage: a missing link 
in sun-climate relationships. J.Atmospheric 
and Solar Terrestrial Physics {\bf 59} 1225.

\item Svensmark H., 2007, Cosmoclimatology: a new theory emerges, 
New Reviews in Astronomy and Geophysics {\bf 48} 18.

\item Svensmark H. and Calder N., 2007,  
The Chilling Stars: A New Theory of 
Climate Change, (published by Icon Books).

\item Taylor F.W., 2005, Elementary Climate Physics, 
ISBN 978 0 19  856733 2. Oxford University Press.

\item Veizer J., 2005, Celestial Climate Driver: a perspective from 
four billion years of the carbon cycle, GeoScience Canada, {\bf 32} 
Number 1, 13. 

\item Weber W., 2010 Strong signature of the active Sun in 100 years 
of terrestrial insolation data, Ann. Phys.(Berlin) No. 6 {\bf 522} 
372-381 /DOI 10.1002/andp.201000019. 

\item Yu F., 2002, Altitude Variation of cosmic ray production of 
aerosols: Implications for global cloudiness and climate, 
Journal of Geophysical Research, {\bf 107} number A7, \\ 
10.1029/2001JA000248. 

\end{enumerate}

\begin{figure}[htb]
\includegraphics[width=40pc]{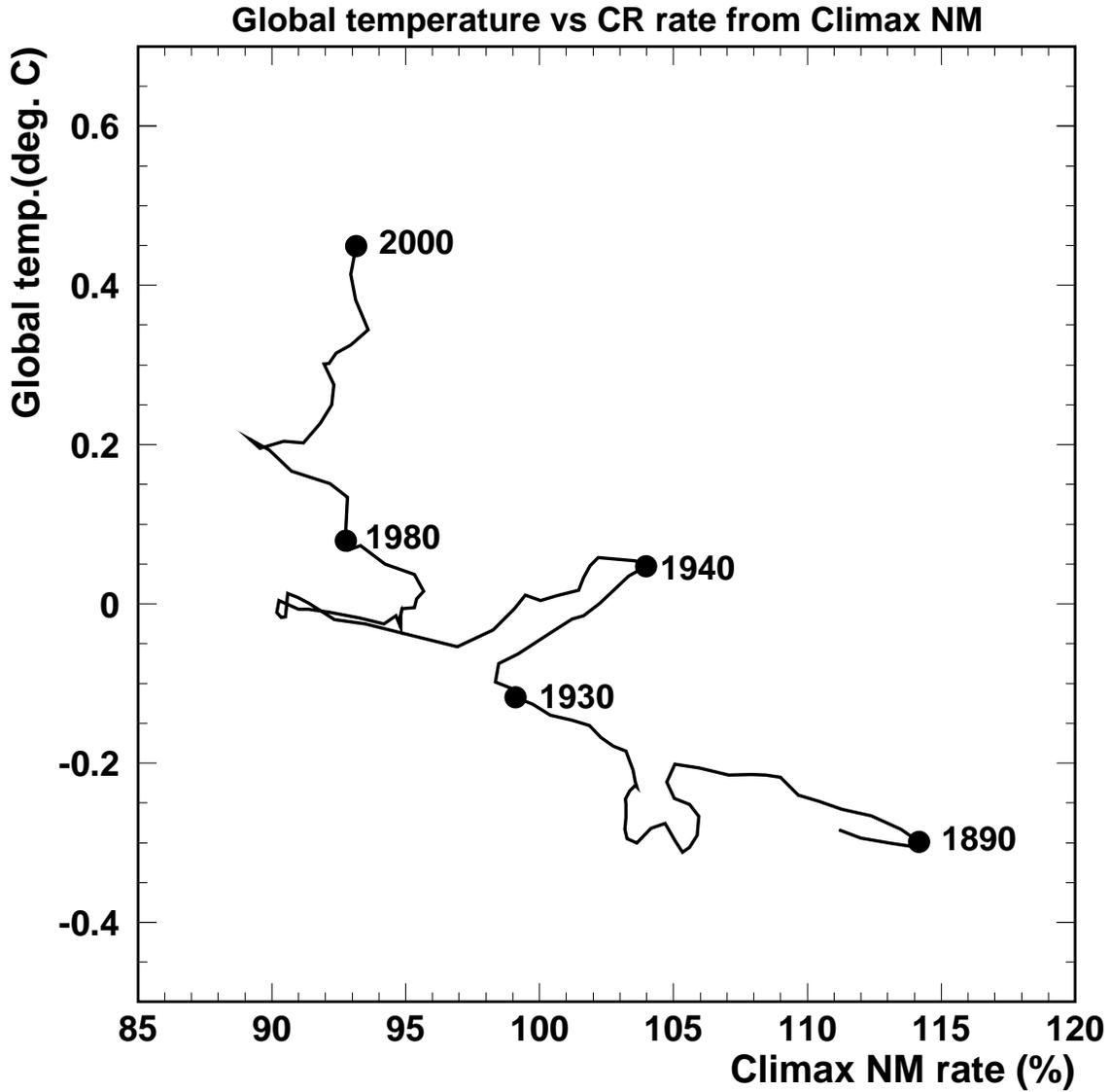}
\caption{\label{fig3} The data from figure \ref{fig2} plotted as equivalent 
Climax neutron monitor rate against temperature anomaly. The solid points 
correspond to the years marked. } 
\end{figure}

\begin{figure}[htb]
\includegraphics[width=40pc]{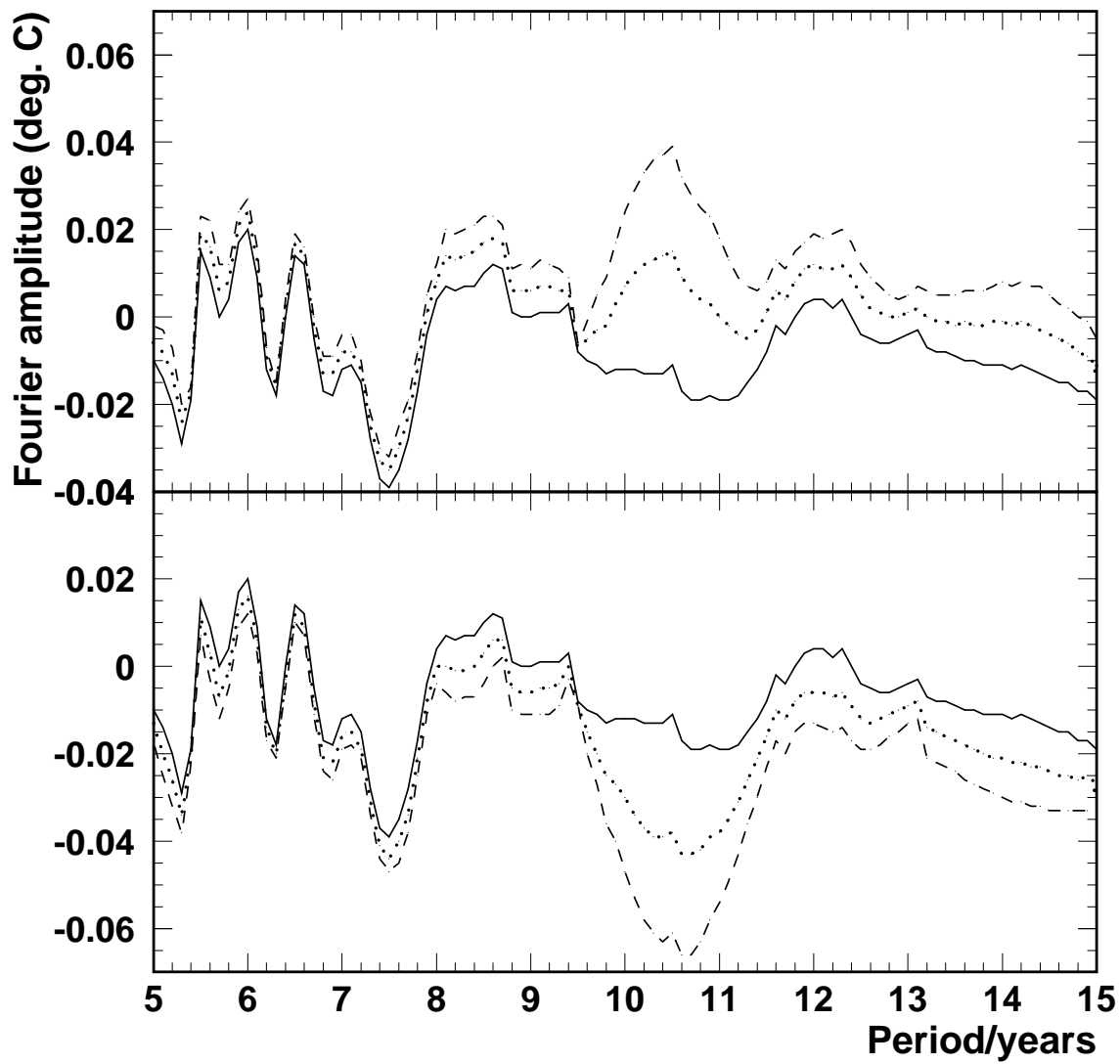}
\caption{\label{fig4} Fourier amplitude components in $^\circ$C
in phase with the equivalent Climax neutron monitor 
wave as a function of wave period. 
The solid curves show the Fourier components of the   
the GISS temperature measurements shown in figure \ref{fig1}.   
The dotted and dashed curves show the results of the 
analysis applied to these  
temperatures with equivalent Climax data normalised to  
temperature waves of amplitude 0.035 and 0.07$^\circ$C, 
respectively, added directly to the temperature data (plus 
in the upper panel, minus in the lower panel). } 
\end{figure}

\begin{figure}[htb]
\includegraphics[width=40pc]{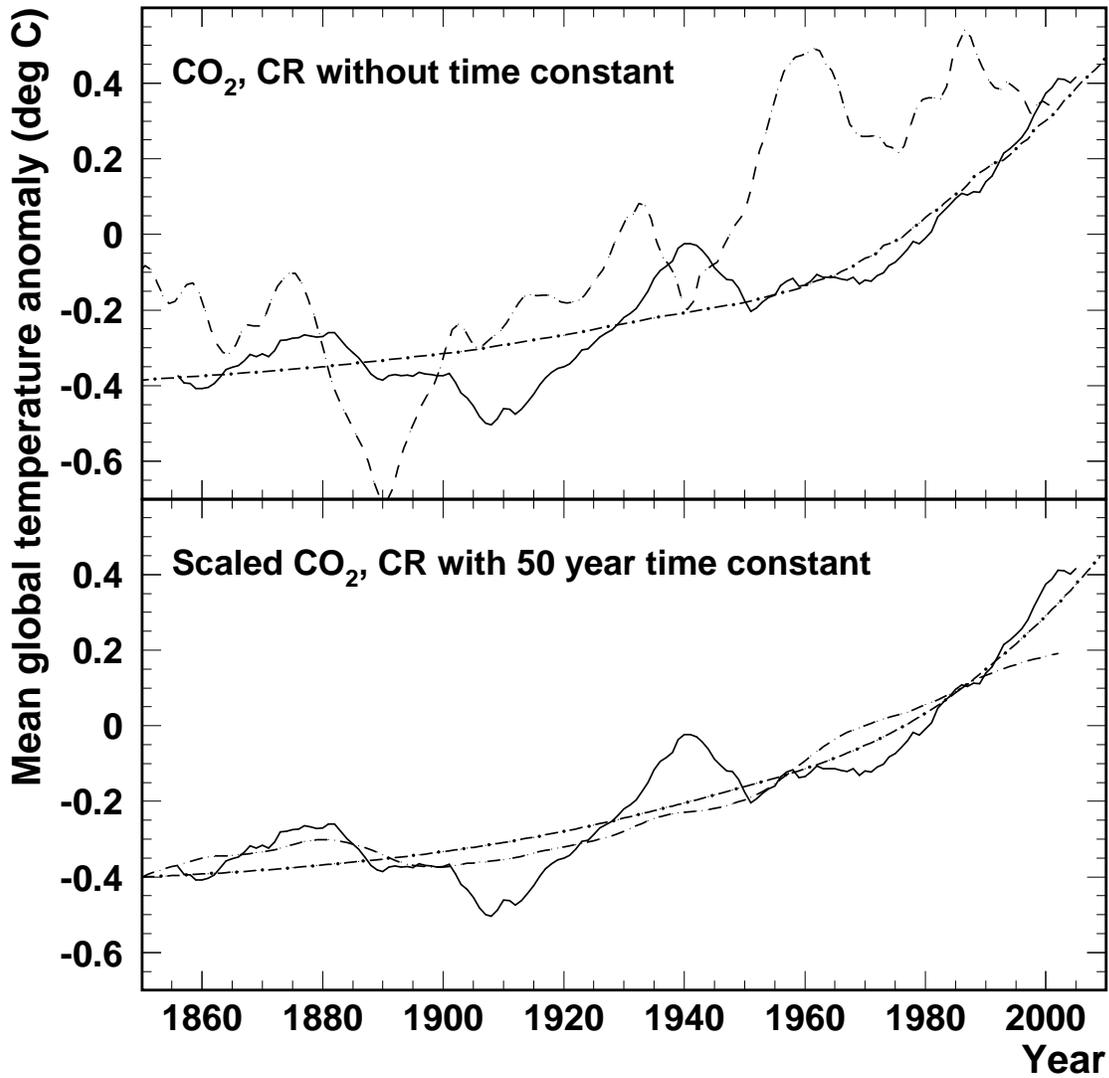}
\caption{\label{fig5} Comparison of the mean global surface temperature 
anomalies (solid 
curves) with the scaled measured CR rates (dashed curves) and CO$_2$ 
concentrations (dash-dotted curves). The CO$_2$ measurements are scaled 
so that $T_{CO2}= k_{CO2}(CO2-285)-0.4$ and the CR measurements so that 
$T_{CR}=k_{CR}(100-CL)$ where $k_{CO2}$ and $k_{CR}$ are normalising 
constants, 
$CO2$ is the concentration of carbon dioxide in the atmosphere in ppmv 
(NOAA 2011) and $CL$ are the Climax rates (McCracken and Beer 2007). 
In the upper plot 
the scaled CO$_2$ and CR measurements were plotted directly with 
$k_{CO2}$=0.0083 and $k_{CR}$=0.05. In the lower plots the CO$_2$ 
and Climax rates were 
passed through an integrating time constant of 50 years. 
The values of the constants to achieve a reasonable representation of 
the data in this case are $k_{CO2}$=0.0175 and $k_{CR}$=0.52.} 
\end{figure}

\end{document}